\documentclass[journal=jacsat,manuscript=article]{achemso}
\usepackage[version=3]{mhchem} 
\usepackage{graphicx}

\author{Henry A. Fernandez}
\email{hf278@exeter.ac.uk}
\author{Freddie Withers}
\author{Saverio Russo}
\author{William L. Barnes}
\email{w.l.barnes@exeter.ac.uk}
\affiliation{Department of Physics and Astronomy, University of Exeter, Exeter, United Kingdom, EX4 4QL}

\title[title]
{Electrically tuneable exciton-polaritons through free electron doping
in monolayer WS$_2$ microcavities
}


\begin{document}
\begin{abstract}
We demonstrate control over light-matter coupling at room temperature combining a field effect transistor (FET) with a tuneable optical microcavity.
Our microcavity FET comprises a monolayer tungsten disulfide WS$_2$ semiconductor which was transferred onto a hexagonal boron nitride flake that acts as a dielectric spacer in the microcavity, and as an electric insulator in the FET.
In our tuneable system, strong coupling between excitons in the monolayer WS$_2$ and cavity photons can be tuned by controlling the cavity length, which we achieved with excellent stability, allowing us to choose from the second to the fifth order of the cavity modes.
Once we achieve the strong coupling regime, we then modify the oscillator strength of excitons in the semiconductor material by modifying the free electron carrier density in the conduction band of the WS$_2$.
This enables strong Coulomb repulsion between free electrons, which reduces the oscillator strength of excitons until the Rabi splitting completely disappears.
We controlled the charge carrier density from 0 up to 3.2 $\times$ 10$^{12}$ cm$^{-2}$, and over this range the Rabi splitting varies from a maximum value that depends on the cavity mode chosen, down to zero, so the system spans the strong to weak coupling regimes.
\end{abstract}

\section{Introduction}
Exciton-polaritons are hybrid bosonic quasiparticles that have properties of both of their constituents: material properties of excitons, responsible for strong non-linear interactions; and optical properties of photons, that make them low mass compared with bare excitons.
These quasiparticles have enabled the observation of cold atom physics including Bose-Einstein condensation on a chip \cite{Kasprzak2006}, superfluidity \cite{Amo2009,Amo2009b,Lerario2017}, and topological polaritons~\cite{Karzig2015}.
At the same time, owing to their hybrid light-matter nature, exciton-polaritons may provide the strong non-linearities needed to deliver the strongly interacting photons \cite{Liang2018} sought by quantum technologies \cite{Reiserer2015}.
This scenario could be significantly enriched by the ability of novel quantum materials to support room temperature operation of exciton-polaritons and the addition of unique functionalities, such as on-chip tuneable strength of the light-matter interaction.\par
Atomically thin semiconductors, part of the family of transition metal dichalcogenides (TMDs), exhibit a range of unique optical properties, which makes them one of the most promising material systems for enabling quantum photonics \cite{Mueller2018}.
Prominent optical transitions in TMDs are widely dominated by excitons with a surprisingly large binding energy \cite{Berkelbach2013} making them viable for room temperature quantum electrodynamics.
Additionally, due to their large exciton oscillator strength, TMDs support strong light-matter interactions even in their atomically thin form \cite{Liu2015,Dufferwiel2015,Wang2016,Flatten2016,Chervy2018,Dufferwiel2018}.
However, the tuneability of light-matter interactions in TMD-based devices, which would enable further control over non-linear interactions, is still in its infancy and further development is needed.\par
To investigate both tuneability and control over the light-matter coupling of exciton-polaritons we designed \cite{Fernandez2017} and fabricated tuneable Fabry-P\'erot microcavities with a TMD-based transistor embedded within the microcavity. We chose the TMD material tungsten disulfide, WS$_2$, which exhibits the largest oscillator strength, and the smallest damping factor among TMDs of chemical composition MX$_2$ (M = W, Mo, X = S, Se) \cite{Li2014}.
These properties make WS$_2$ the optimum semiconductor TMD for strong coupling measurements of exciton-polaritons at visible frequencies, at room temperature, and in microcavities of moderate quality factor, $Q \approx 100$.
In our device, we increase free carrier density of a WS$_2$ monolayer, which produces a strong Coulomb screening of the free electrons that populate the conduction band~\cite{Chernikov2015}. This screening reduces the oscillator strength of neutral excitons, which therefore weakens the coupling strength of exciton-polaritons \cite{Houdre1995}.\par
Electrical control of light-matter coupling has been demonstrated in a monolythic microcavity filled with carbon nanotubes \cite{Gao2018}, where near infrared excitons were strongly coupled to a single microcavity mode.
By applying a range of gate voltages the system was observed to cross reversibly from the weak to the strong coupling regime.
Recently, this phenomenon was observed in the visible range using a monolythic TMD microcavity \cite{Chakraborty2018}, showing similar results.
In our study we significantly extend the work by applying gate voltages sufficient to reach the saturation regime of the light-matter coupling strength, characterized by the vacuum Rabi splitting.
For sufficiently negative values of the gate voltage, the Rabi splitting saturates to its maximum value, whilst for sufficiently positive values of the gate voltage (high electron carrier density), the Rabi splitting was completely eliminated.
These observations allow us to explain the functional dependence of the Rabi splitting on the gate voltage, and to probe the nature of the Coulomb screening process.\par
In this report we studied a WS$_2$ microcavity in which the TMD excitons are strongly coupled to microcavity modes that can be chosen from the third-order to the fifth-order by controlling the cavity length in a tuneable way.
We observed a dependence of the Rabi splitting on the electrically-controlled density of free electrons, a density that ranges from 0 up to 3$\times$10$^{12}$ $cm^{-2}$. 
Over this range the Rabi splitting varies from a maximum value (which depends on the order of the chosen cavity mode) down to zero so that the hybrid system spans the strong to weak coupling regimes.
\section{Results}
\subsection{WS$_2$-based field effect transistor}
\begin{figure}[h!]
    \centering
    \includegraphics[width=0.8\textwidth]{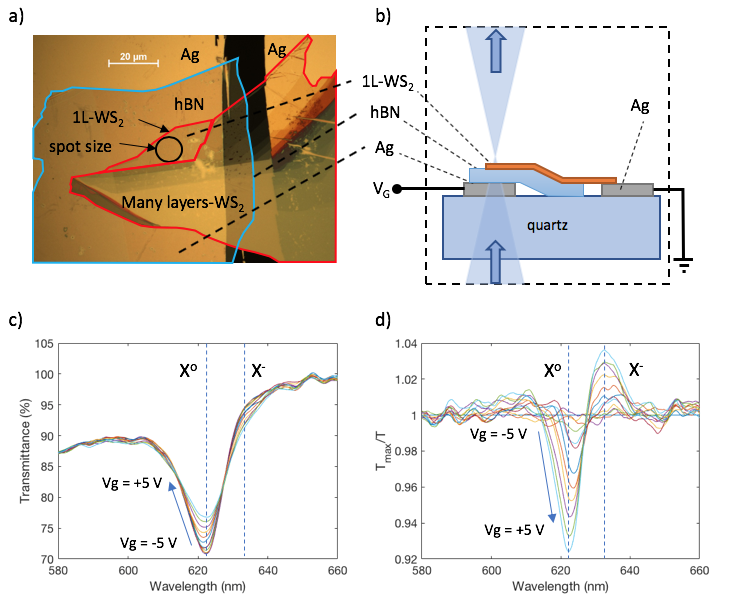}
    \caption{(a) Top view microscope image of the WS$_2$/hBN/Ag heterostructure on quartz.
    Colour lines indicate flake edges: blue for hBN, and red for WS$_2$.
    (b) Schematic cross-sections of the WS$_2$-based transistor.
    Arrows indicate the propagation direction of the white light.
    (c) White light transmittance of the 1L-WS$_2$ for different gate voltages from -5 V up to +5 V.
    The positions of neutral excitons X$^o$ and negatively charged excitons, or negative trions X$^-$, are indicated.
    (d) Data shown in (c) normalized to the transmittance spectrum for V$_G$ = -5 V, which is the reference T$_{max}$.
    This plot shows how the transmission for both the neutral excitons and the negative trions change for the different values of gate voltage.}
    \label{fig:fig1}
\end{figure}

We first studied the WS$_2$-based field effect transistor without a microcavity.
The transistor consisted of a Van der Waals heterostructure composed by WS$_2$ placed on top of a hexagonal boron nitride (hBN) flake (more details in the experimental section).
The WS$_2$/hBN heterostructure was placed on a silver (Ag) film, which acts as a microcavity mirror (bottom mirror) as well as the gate contact of the transistor while WS$_2$ was kept as a ground contact (see figure \ref{fig:fig1}(a) and (b)).
The hBN flake plays an important role in both the optical and the electronic performance of the transistor. For the optics, hBN acts as a dielectric spacer; the thickness of hBN determines the position at which the WS$_2$ flake is placed in the microcavity, therefore the confined field intensity which TMD excitons interact with (see supporting information figure S1). For electronics, hBN acts as a dielectric insulator ($\epsilon \sim 3-4$)\cite{Zhang2017}, which prevents electrons to leak between the WS$_2$ and the Ag gate electrode.
Transmission measurements were performed through this device and we studied the evolution of the minimum in  transmission associated with the WS$_2$ exciton for different values of the gate voltage.
Figure \ref{fig:fig1}(a) shows a microscope image of the top view of the WS$_2$ transistor; coloured lines are drawn to indicate the edges of the different materials of the heterostructure, red for WS$_2$, and blue for hBN. The spot size for light collection is also indicated on the figure, which has a diameter of 10 $\mu m$.
The area of interest is a monolayer WS$_2$ (1L-WS$_2$) electrically in contact with a multilayer flake (see figure \ref{fig:fig1}(a)) allowing for a minimized contact barrier height between the monolayer and the multilayer TMD.
The gate voltage is applied to the WS$_2$ while keeping the Ag electrode to ground.
In figure \ref{fig:fig1}(b) a schematic side view of the sample is shown, in which the direction of propagation of the transmitted light is indicated by arrows, and the position of the electrical contacts is shown.
We collected white light transmission spectra for different values of the gate voltage V$_G$, as shown in figure \ref{fig:fig1}(c).
In this figure a transmittance minimum is observed at a wavelength of 622 nm, and is associated with the neutral exciton transition of the 1L-WS$_2$ labelled X$^o$.
In this figure we also observe the appearance of a weak transmittance dip at a wavelength of 632 nm, which we associate with a negatively charged exciton, a negative trion, which is labelled as X$^-$.
We observe a weakening in the transmission minimum associated with the neutral excitons as we sweep the gate voltage from -5 V to +5 V.
We also observed that the feature associated with the negative trions strengthens.
To clarify this observation we extract the change in the transmission associated with X$^o$ and X$^-$ as a function of V$_G$. These data are shown in figure \ref{fig:fig1}(d) where values below one indicate how much the transmission is reduced in strength, whilst values above one indicate how much the transmission increases in strength.
We observed no significant electrical current through the hBN barrier in this range of voltages (see supporting information figure S5), which suggests that the free electrons are accumulated on the WS$_2$ which then leads to a reduction in the excitation rate of neutral excitons due to strong Coulomb repulsion between free carriers filling the conduction band \cite{Zhu2015}.
\subsection{Tuneable microcavity}
\begin{figure}
    \centering
    \includegraphics[width=0.8\textwidth]{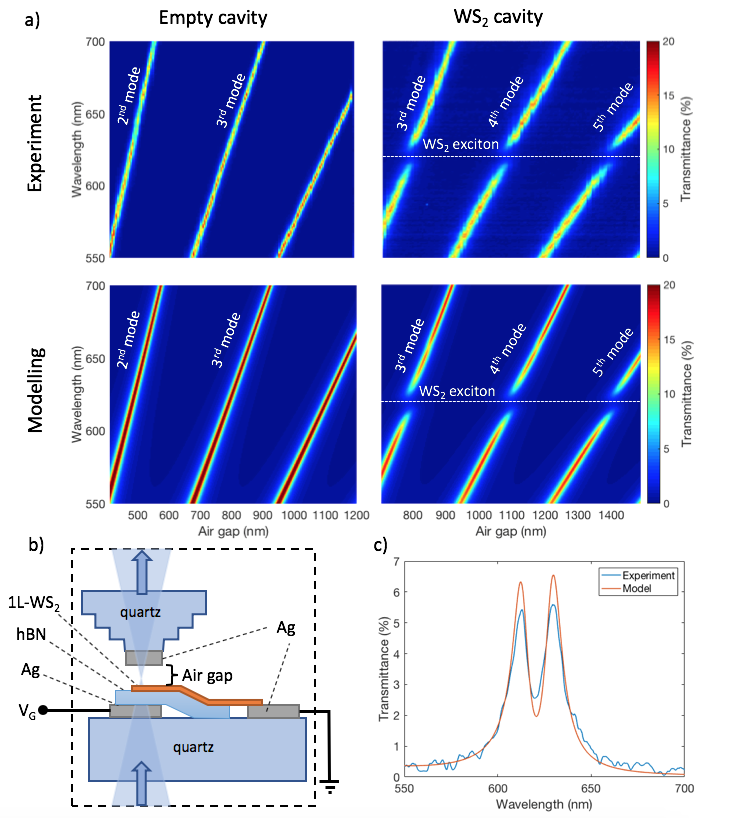}
    \caption{(a) Experimental data, and modelling of the empty cavity modes, and the WS$_2$-cavity modes as a function of the air gap between the WS$_2$ flake and the top silver mirror.
    (b) Schematic of the tuneable microcavity, which consists of the same sample shown in figure \ref{fig:fig1}(b) but with a top mirror added leaving a small gap between the bottom and top mirrors.
    (c) Experimental and calculated transmittance of a strongly coupled WS$_2$ microcavity.
    Strong coupling was achieved for a cavity air gap of 787 $\pm$ 3 nm, which corresponds to the third-order cavity mode.}
    \label{fig:fig2}
\end{figure}

A tuneable microcavity was completed by adding a second silver mirror (top mirror) close to the top of the WS$_2$ transistor, as shown in figure \ref{fig:fig2}(b).
For the microcavity we collected transmission spectra as a function of the thickness of the air gap between the hBN spacer and the top silver mirror in a region where there is no WS$_2$. 
These transmission spectra are then compared to the modelled spectra from which we extract the experimental value for the thickness of the air gap (for details on modelling see the supporting information),
as shown in the left panels of figure \ref{fig:fig2}(a). Here we identify the second-, third-, and fourth-order cavity modes.
We observe excellent stability of our tuneable system, and very good quality of the microcavity modes compared to modelling.
We repeated these measurements in a region of the microcavity where the 1L-WS$_2$ was present, keeping the gate voltage fixed at 0 V.
These measurements were compared to the model and are shown in the right panels of \ref{fig:fig2}(a).
Here we observe a splitting of the cavity modes at a wavelength of 622 nm, associated with the neutral exciton transition of WS$_2$.
The mode splitting of the three cavity modes observed is greater than that of both the empty cavity mode and the exciton line widths (36 and 35 meV respectively), demonstrating that the system is in the strong coupling regime.
We fit the calculated transmittance of a strongly coupled 1L-WS$_2$-microcavity with an air gap of 787 nm, which corresponds to the third-order cavity mode, to the experimental observation as shown in figure \ref{fig:fig2}(c), where the free parameter was the oscillator strength of WS$_2$; we found that the best fit was achieved for an oscillator strength of $f = 2.6$.
A comparison between the calculated and the measured transmittance spectra of a bare 1L-WS$_2$ is shown in the supporting information figure S3.
The oscillator strength estimated in our experiments are significantly larger than values reported for WS$_2$ on SiO$_2$ or Al$_2$O$_3$ substrates\cite{Wang2016,Chakraborty2018}.
The origin of this discrepancy could be the difference in the dielectric environment provided by the hBN flake in our devices, and also to the fact that hBN provides an atomically smooth surface (roughness lower than 0.5 nm as measured by atomic force microscopy) which may reduce surface scattering and trapping effects in the WS$_2$/hBN interface.
\subsection{Electrical control of the light-matter coupling}
\begin{figure}[h!]
    \centering
    \includegraphics[width=0.8\textwidth]{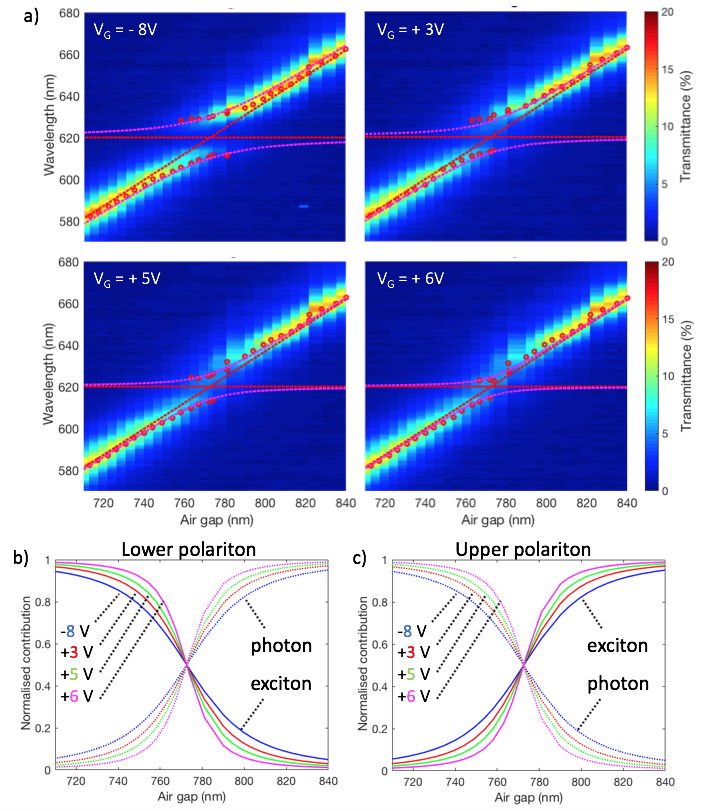}
    \caption{(a) Analysis of the transmittance spectra of the third-order cavity mode of the WS$_2$ microcavity for different values of the gate voltage.
    Dashed oblique lines are the uncoupled cavity mode obtained from modelling an empty microcavity (see figure \ref{fig:fig2}(a)), horizontal dashed lines indicate the neutral exciton transition of WS$_2$, and curved dashed lines are the lower and upper polariton bands obtained from the coupled oscillator model.
    Circles are the experimental peak center of the transmission data.
    (b), and (c) are the Hopfield coefficients of the lower and upper polariton bands respectively.
    The different colours correspond to different values of the gate voltage.
    Continuous lines correspond to the exciton contribution to the polariton bands, and dashed lines correspond to the cavity photon contribution.}
    \label{fig:fig3}
\end{figure}

\begin{figure}[h!]
    \centering
    \includegraphics[width=0.4\textwidth]{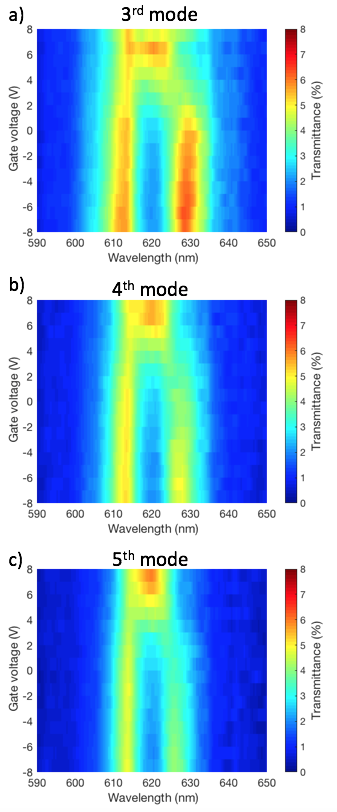}
    \caption{Transmittance spectra of a WS$_2$ microcavity for different values of the gate voltage, and for a fixed air gap: (a) 787 $\pm$ 3 nm, (b) 1097 $\pm$ 3 nm, and (c) 1408 $\pm$ 3 nm, which correspond to the third, fourth and fifth cavity modes respectively.}
    \label{fig:fig4}
\end{figure}

We next discuss the evolution of the cavity mode splitting in the strong coupling regime for different gate voltages ranging from -8 V up to +8 V.
To study the exciton and photon nature of the polariton bands we performed a two coupled oscillator model analysis of the experimental data, from which we found the Hopfield coefficients as described in the supporting information.
We focused this analysis on the splitting of the third-order cavity mode, for four values of the gate voltage: -8 V, +3V, +5 V, and +6 V.
These results are shown in figure \ref{fig:fig3}(a), in which the four values of the gate voltage are indicated.
On each plot we show the position of the neutral exciton transition of WS$_2$ as a horizontal dashed line, and the third-order cavity mode as an oblique dashed line.
The circles are the peak maxima obtained by fitting Pseudo-Voigt functions to the experimental data, for each value of the air gap.
We then fit the coupled oscillator model to the position of the peak maxima.
The input parameter in this analysis is the experimental value of the Rabi splitting, which corresponds to the transmission peak splitting at the point where the neutral exciton transition crosses the bare cavity mode.
From the coupled oscillator model we find the Hopfield coefficients which we show in figure \ref{fig:fig3}(b) for the lower polariton band, and (c) for the upper polariton band.
We observe that the lower polariton band becomes more photon-like nature for the most positive values of the gate voltage, and therefore less excitonic nature.
The upper polariton band becomes more excitonic nature for the most positive values of the gate voltage.
The behaviour of both of the polariton bands suggests a decoupling between excitons and
cavity photons that occurs when a positive gate voltage is applied.
A decoupling between excitons and cavity photons can also be observed as a reduction of the Rabi splitting
as a function of the gate voltage, as shown in figure \ref{fig:fig4}.
In this figure we show the transmittance peak splitting for (a) the third, (b) the fourth, and (c) the fifth-order cavity modes as a function of the gate voltage.
These modes correspond to fixed air gaps, being 787 $\pm$ 3 nm, 1097 $\pm$ 3 nm, and 1408 $\pm$ 3 nm 
respectively.
For the three modes we observe a dependence of the Rabi splitting on the gate voltage; the Rabi splitting reaches a saturation value when sufficiently negative voltages are applied to the transistor.
In addition the maximum splitting is lower for higher order cavity modes, as expected due to the weaker fields associated with the higher-order modes (see also supporting information figure S6).
From these measurements we obtain the Rabi splitting as a function of the gate voltage for the three cavity modes as shown in figure \ref{fig:fig5}(a).
In this figure we also indicate the exciton line width, which is equal or larger to any of the three cavity mode widths, indicating that the system is in the strong coupling regime when the Rabi splitting is greater than the exciton line width.
\subsection{Rabi splitting and density of free electrons}
\begin{figure}
    \centering
    \includegraphics[width=0.4\textwidth]{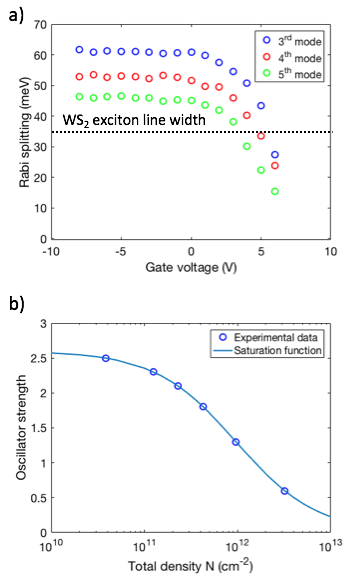}
    \caption{(a) Rabi splitting as a function of the gate voltage obtained from the data shown in figure \ref{fig:fig4}.
    The three different modes are shown as different colours: blue for 3$^{rd}$ mode, red for 4$^{th}$ mode, and green for the 5$^{th}$ mode.
    (b) Oscillator strength as a function of the total density of electrons $N$. The saturation function (eq.\ref{eqn:saturation1}) is also shown.}
    \label{fig:fig5}
\end{figure}

We now focus our attention on obtaining the dependence of the oscillator strength of WS$_2$ as we change the gate voltage, a dependence that we can relate to an increase of the density of free electrons in the WS$_2$.
To do this we analyse the splitting of the third-order cavity mode.
We first calculate the transmittance spectrum as shown in figure \ref{fig:fig2}(c) with an oscillator strength of $f_0 = 2.6$, which results in a Rabi splitting of 60 meV that corresponds to the splitting of the third-order mode for a gate voltage of 0 V.
Prior to this measurements, we applied a sufficiently positive gate voltage to remove any excess of free electrons in WS$_2$, so $f_0$ is our initial oscillator strength for zero density of free electrons in the TMD.
Next we vary the oscillator strength in our calculations to reproduce the values of the Rabi splitting shown in figure \ref{fig:fig5}(a) for the different values of the gate voltage.
The density of free electrons $N$ can be obtained by inserting the known values of the oscillator strength, $f$, in equation (\ref{eqn:saturation1}),
\begin{equation}
    f(N) = \frac{f_0}{1+\frac{N}{N_s}},
    \label{eqn:saturation1}
\end{equation}
where the saturation number $N_s$ is defined as the number density of carriers at which the oscillator strength is reduced to a half of the initial value $f_0$.
The value of $N_s$ can be estimated for any 2-dimensional system that allows for the excitation of excitons and free carriers either in the conduction or the valence band.
For our analysis we have used $N_s =1/8.3\pi a_{EB}$ extracted from \cite{Schmitt-Rink1985}, where $a_{EB}$ is the effective exciton-Bohr radius of WS$_2$, which has been taken from reference~\citenum{Ye2014} to be a value of 2 nm corresponding to the X$^0$ exciton transition.\par
These results are shown in figure \ref{fig:fig5}(b), where we also include the saturation function as described in equation (\ref{eqn:saturation1}).
The density of free electrons $N$ found in this way agrees well with what is expected in a TMD system that behaves below the level of Mott transition  \cite{Chernikov2015,Chernikov2015b}, which has been estimated to occur at a density above 10$^{13}$ $cm^{-2}$.
In our system, we found the saturation density to be $9.6\times 10^{11}\ cm^{-2}$, and in the range of voltages applied, the density of free electrons in the WS$_2$ ranges from 0 up to $3.2\times 10^{12}\ cm^{-2}$.
\section{Summary and conclusions}
In summary, we studied continuous control over the light-matter coupling of exciton-polaritons in a WS$_2$ microcavity at room temperature.
We have achieved this by controlling the charge carrier density in the semiconductor material.
Due to the Coulomb interaction the free carriers repel each other and screen the oscillator strength of excitons in WS$_2$.
A specially designed WS$_2$ based field effect transistor was built on one of the mirrors of the microcavity \cite{Fernandez2017}, then a second mirror was placed on top of the transistor leaving an air gap which we were able to control with a precision of $\pm$3 nm in our tuneable system.
This tuneability allowed us to study the polariton bands as a function of the thickness of the air gap, and we accessed different cavity modes ranging from the second-order to the fifth-order.
For each of the cavity modes, at the resonant condition with WS$_2$ excitons forming exciton-polaritons, we were able to record transmittance spectra over a range of gate voltages, from -8 V up to +8 V, observing a dependence of the Rabi splitting as a function of the gate voltage.
Over this range of voltages we calculated the variation of the oscillator strength in the TMD film, extracted from modelling the transmittance spectra to match the measured values of the Rabi splitting.
We attributed the change in the oscillator strength of WS$_2$ to Coulomb screening by the free electrons populating the conduction band of the TMD, we then were able to extract the density of free carriers in the conduction band, which ranges from 0 cm$^{-2}$, at a gate voltage of 0 V, up to 3.2 $\times$ 10$^{12}$ cm$^{-2}$, for a gate voltage of +8 V.
Over this range of density of free carriers, the Rabi splitting ranges from a maximum value, which depends on the cavity mode chosen, down to zero, so the system spans the strong to weak coupling regimes.
This change in the Rabi splitting could also be extended to p-type semiconductors with a strong excitonic resonance by injecting holes in the valence band of the material.
\section{Experimental section}
\subsection{Heterostructure}
WS$_2$ flakes where exfoliated from single crystals purchased from HQgraphene, hBN flakes where obtained through the same method.
hBN single crystals were purchased from Manchester Nanomaterials.
The flakes where then transferred onto silver films deposited on quartz glass (bottom mirror) by thermal evaporation.
The thickness of the silver films was nominally 40 nm, later confirmed by AFM.
\subsection{Tuneable microcavity setup}
The tuneable microcavity was completed by a top mirror placed on a piezo-electric stage (Thorlabs MAX313D)
controlled externally (Thorlabs MDT30B) through LabView. Both mirrors had five degrees
of freedom, three translations and two rotations. The alignment procedure to make both of the mirrors parallel consisted of sending red light by using a band pass filter in front of the white light source, we then examined the transmitted light on a digital camera, the transmitted light consisted of an interference pattern of rings.
Then, by rotating the top mirror we adjusted the ring pattern until we reached the centre of the pattern.
The accuracy of this procedure depended on the magnification of the camera, which in turn determined how many lines of the ring pattern could be observed on the computer screen, the accuracy also depended on the quality of the microcavity and the line-width of the band-pass filter, which determined the width of the lines that made up the ring pattern.
\subsection{Optical and electrical measurements}
Transmission measurements where performed in a confocal setup.
The light source (Thorlabs\textsuperscript{\textregistered} SLS201L/M) was fibre coupled, collimated and focused by x10 objective lenses through the bottom mirror. To collect the transmitted light through the top mirror a x50 objective lens was used (Olympus\textsuperscript{\textregistered} SLMPLN50X), which was then focused on a fibre coupled spectrometer (OceanOptics\textsuperscript{\textregistered} Flame).
The gate voltage was applied by a Keithley\textsuperscript{\textregistered} 2450 SMU; gate voltage, and tunnelling current were recorded for the different values of the cavity length. The top mirror was placed on a piezo-stage  (Thorlabs\textsuperscript{\textregistered} MAX312D, MDT630B) which allows for a control of the cavity length with a precision of $\pm$ 3 nm.
Transmittance, piezo-stage voltage, and gate voltage measurements were synchronised on LabView\textsuperscript{\textregistered}.

\begin{acknowledgement}

The authors acknowledge Prof. Misha Portnoy for several insightful discussions. HF, and WLB thank Dr. Lucas Flatten for his advice on the tuneable microcavity setup.
The authors also acknowledge financial support from the Engineering and Physical Sciences Research Council (EPSRC) of the United Kingdom, via the EPSRC Centre for Doctoral Training in Metamaterials (Grant No. EP/L015331/1).
FW acknwoledges support from the Royal Academy of Engineering. SR acknowledges financial support from The Leverhulme Trust, research grants ``Quantum Drums" and ``Quantum Revolution".
WLB acknowledges the support of the European Research Council through the project Photmat (ERC-2016-ADG-742222: www.photmat.eu).

\end{acknowledgement}

\begin{suppinfo}
Calculations of the electric field in the microcavity system, and characterization of the different materials can be found in the supporting information.
\end{suppinfo}

\bibliography{bibliography}

\providecommand{\latin}[1]{#1}
\makeatletter
\providecommand{\doi}
  {\begingroup\let\do\@makeother\dospecials
  \catcode`\{=1 \catcode`\}=2 \doi@aux}
\providecommand{\doi@aux}[1]{\endgroup\texttt{#1}}
\makeatother
\providecommand*\mcitethebibliography{\thebibliography}
\csname @ifundefined\endcsname{endmcitethebibliography}
  {\let\endmcitethebibliography\endthebibliography}{}
\begin{mcitethebibliography}{27}
\providecommand*\natexlab[1]{#1}
\providecommand*\mciteSetBstSublistMode[1]{}
\providecommand*\mciteSetBstMaxWidthForm[2]{}
\providecommand*\mciteBstWouldAddEndPuncttrue
  {\def\EndOfBibitem{\unskip.}}
\providecommand*\mciteBstWouldAddEndPunctfalse
  {\let\EndOfBibitem\relax}
\providecommand*\mciteSetBstMidEndSepPunct[3]{}
\providecommand*\mciteSetBstSublistLabelBeginEnd[3]{}
\providecommand*\EndOfBibitem{}
\mciteSetBstSublistMode{f}
\mciteSetBstMaxWidthForm{subitem}{(\alph{mcitesubitemcount})}
\mciteSetBstSublistLabelBeginEnd
  {\mcitemaxwidthsubitemform\space}
  {\relax}
  {\relax}

\bibitem[Kasprzak \latin{et~al.}(2006)Kasprzak, Richard, Kundermann, Baas,
  Jeambrun, Keeling, Marchetti, Szyma{\'{n}}ska, Andr{\'{e}}, Staehli, Savona,
  Littlewood, Deveaud, and Dang]{Kasprzak2006}
Kasprzak,~J.; Richard,~M.; Kundermann,~S.; Baas,~a.; Jeambrun,~P.; Keeling,~J.
  M.~J.; Marchetti,~F.~M.; Szyma{\'{n}}ska,~M.~H.; Andr{\'{e}},~R.;
  Staehli,~J.~L.; Savona,~V.; Littlewood,~P.~B.; Deveaud,~B.; Dang,~L.~S.
  \emph{Nature} \textbf{2006}, \emph{443}, 409--414\relax
\mciteBstWouldAddEndPuncttrue
\mciteSetBstMidEndSepPunct{\mcitedefaultmidpunct}
{\mcitedefaultendpunct}{\mcitedefaultseppunct}\relax
\EndOfBibitem
\bibitem[Amo \latin{et~al.}(2009)Amo, Sanvitto, Laussy, Ballarini, del Valle,
  Martin, Lema{\^{i}}tre, Bloch, Krizhanovskii, Skolnick, Tejedor, and
  Vi{\~{n}}a]{Amo2009}
Amo,~A.; Sanvitto,~D.; Laussy,~F.~P.; Ballarini,~D.; del Valle,~E.;
  Martin,~M.~D.; Lema{\^{i}}tre,~A.; Bloch,~J.; Krizhanovskii,~D.~N.;
  Skolnick,~M.~S.; Tejedor,~C.; Vi{\~{n}}a,~L. \emph{Nature} \textbf{2009},
  \emph{457}, 291\relax
\mciteBstWouldAddEndPuncttrue
\mciteSetBstMidEndSepPunct{\mcitedefaultmidpunct}
{\mcitedefaultendpunct}{\mcitedefaultseppunct}\relax
\EndOfBibitem
\bibitem[Amo \latin{et~al.}(2009)Amo, Lefr{\`{e}}re, Pigeon, Adrados, Ciuti,
  Carusotto, Houdr{\'{e}}, Giacobino, and Bramati]{Amo2009b}
Amo,~A.; Lefr{\`{e}}re,~J.; Pigeon,~S.; Adrados,~C.; Ciuti,~C.; Carusotto,~I.;
  Houdr{\'{e}},~R.; Giacobino,~E.; Bramati,~A. \emph{Nature Physics}
  \textbf{2009}, \emph{5}, 805\relax
\mciteBstWouldAddEndPuncttrue
\mciteSetBstMidEndSepPunct{\mcitedefaultmidpunct}
{\mcitedefaultendpunct}{\mcitedefaultseppunct}\relax
\EndOfBibitem
\bibitem[Lerario \latin{et~al.}(2017)Lerario, Fieramosca, Barachati, Ballarini,
  Daskalakis, Dominici, {De Giorgi}, Maier, Gigli, K{\'{e}}na-Cohen, and
  Sanvitto]{Lerario2017}
Lerario,~G.; Fieramosca,~A.; Barachati,~F.; Ballarini,~D.; Daskalakis,~K.~S.;
  Dominici,~L.; {De Giorgi},~M.; Maier,~S.~A.; Gigli,~G.; K{\'{e}}na-Cohen,~S.;
  Sanvitto,~D. \emph{Nature Physics} \textbf{2017}, \emph{13}, 837\relax
\mciteBstWouldAddEndPuncttrue
\mciteSetBstMidEndSepPunct{\mcitedefaultmidpunct}
{\mcitedefaultendpunct}{\mcitedefaultseppunct}\relax
\EndOfBibitem
\bibitem[Karzig \latin{et~al.}(2015)Karzig, Bardyn, Lindner, and
  Refael]{Karzig2015}
Karzig,~T.; Bardyn,~C.-E.; Lindner,~N.~H.; Refael,~G. \emph{Phys. Rev. X}
  \textbf{2015}, \emph{5}, 031001\relax
\mciteBstWouldAddEndPuncttrue
\mciteSetBstMidEndSepPunct{\mcitedefaultmidpunct}
{\mcitedefaultendpunct}{\mcitedefaultseppunct}\relax
\EndOfBibitem
\bibitem[Liang \latin{et~al.}(2018)Liang, Venkatramani, Cantu, Nicholson,
  Gullans, Gorshkov, Thompson, Chin, Lukin, and Vuleti{\'c}]{Liang2018}
Liang,~Q.-Y.; Venkatramani,~A.~V.; Cantu,~S.~H.; Nicholson,~T.~L.;
  Gullans,~M.~J.; Gorshkov,~A.~V.; Thompson,~J.~D.; Chin,~C.; Lukin,~M.~D.;
  Vuleti{\'c},~V. \emph{Science} \textbf{2018}, \emph{359}, 783--786\relax
\mciteBstWouldAddEndPuncttrue
\mciteSetBstMidEndSepPunct{\mcitedefaultmidpunct}
{\mcitedefaultendpunct}{\mcitedefaultseppunct}\relax
\EndOfBibitem
\bibitem[Reiserer and Rempe(2015)Reiserer, and Rempe]{Reiserer2015}
Reiserer,~A.; Rempe,~G. \emph{Rev. Mod. Phys.} \textbf{2015}, \emph{87},
  1379--1418\relax
\mciteBstWouldAddEndPuncttrue
\mciteSetBstMidEndSepPunct{\mcitedefaultmidpunct}
{\mcitedefaultendpunct}{\mcitedefaultseppunct}\relax
\EndOfBibitem
\bibitem[Mueller and Malic(2018)Mueller, and Malic]{Mueller2018}
Mueller,~T.; Malic,~E. \emph{npj 2D Materials and Applications} \textbf{2018},
  \emph{2}, 29\relax
\mciteBstWouldAddEndPuncttrue
\mciteSetBstMidEndSepPunct{\mcitedefaultmidpunct}
{\mcitedefaultendpunct}{\mcitedefaultseppunct}\relax
\EndOfBibitem
\bibitem[Berkelbach \latin{et~al.}(2013)Berkelbach, Hybertsen, and
  Reichman]{Berkelbach2013}
Berkelbach,~T.~C.; Hybertsen,~M.~S.; Reichman,~D.~R. \emph{Physical Review B}
  \textbf{2013}, \emph{88}, 045318\relax
\mciteBstWouldAddEndPuncttrue
\mciteSetBstMidEndSepPunct{\mcitedefaultmidpunct}
{\mcitedefaultendpunct}{\mcitedefaultseppunct}\relax
\EndOfBibitem
\bibitem[Liu \latin{et~al.}(2015)Liu, Galfsky, Sun, Xia, Lin, Lee,
  K{\'{e}}na-Cohen, and Menon]{Liu2015}
Liu,~X.; Galfsky,~T.; Sun,~Z.; Xia,~F.; Lin,~E.-c.; Lee,~Y.-H.;
  K{\'{e}}na-Cohen,~S.; Menon,~V.~M. \emph{Nature Photonics} \textbf{2015},
  \emph{9}, 30--34\relax
\mciteBstWouldAddEndPuncttrue
\mciteSetBstMidEndSepPunct{\mcitedefaultmidpunct}
{\mcitedefaultendpunct}{\mcitedefaultseppunct}\relax
\EndOfBibitem
\bibitem[Dufferwiel \latin{et~al.}(2015)Dufferwiel, Schwarz, Withers, Trichet,
  Li, Sich, {Del Pozo-Zamudio}, Clark, Nalitov, Solnyshkov, Malpuech,
  Novoselov, Smith, Skolnick, Krizhanovskii, and Tartakovskii]{Dufferwiel2015}
Dufferwiel,~S. \latin{et~al.}  \emph{Nature Communications} \textbf{2015},
  \emph{6}, 8579\relax
\mciteBstWouldAddEndPuncttrue
\mciteSetBstMidEndSepPunct{\mcitedefaultmidpunct}
{\mcitedefaultendpunct}{\mcitedefaultseppunct}\relax
\EndOfBibitem
\bibitem[Wang \latin{et~al.}(2016)Wang, Li, Chervy, Shalabney, Azzini, Orgiu,
  Hutchison, Genet, Samor{\`{i}}, and Ebbesen]{Wang2016}
Wang,~S.; Li,~S.; Chervy,~T.; Shalabney,~A.; Azzini,~S.; Orgiu,~E.;
  Hutchison,~J.~A.; Genet,~C.; Samor{\`{i}},~P.; Ebbesen,~T.~W. \emph{Nano
  Letters} \textbf{2016}, \emph{16}, 4368--4374\relax
\mciteBstWouldAddEndPuncttrue
\mciteSetBstMidEndSepPunct{\mcitedefaultmidpunct}
{\mcitedefaultendpunct}{\mcitedefaultseppunct}\relax
\EndOfBibitem
\bibitem[Flatten \latin{et~al.}(2016)Flatten, He, Coles, Trichet, Powell,
  Taylor, Warner, and Smith]{Flatten2016}
Flatten,~L.~C.; He,~Z.; Coles,~D.~M.; Trichet,~A.~A.; Powell,~A.~W.;
  Taylor,~R.~A.; Warner,~J.~H.; Smith,~J.~M. \emph{Scientific Reports}
  \textbf{2016}, \emph{6}, 1--7\relax
\mciteBstWouldAddEndPuncttrue
\mciteSetBstMidEndSepPunct{\mcitedefaultmidpunct}
{\mcitedefaultendpunct}{\mcitedefaultseppunct}\relax
\EndOfBibitem
\bibitem[Chervy \latin{et~al.}(2018)Chervy, Azzini, Lorchat, Wang, Gorodetski,
  Hutchison, Berciaud, Ebbesen, and Genet]{Chervy2018}
Chervy,~T.; Azzini,~S.; Lorchat,~E.; Wang,~S.; Gorodetski,~Y.;
  Hutchison,~J.~A.; Berciaud,~S.; Ebbesen,~T.~W.; Genet,~C. \emph{ACS
  Photonics} \textbf{2018}, \emph{5}, 1281--1287\relax
\mciteBstWouldAddEndPuncttrue
\mciteSetBstMidEndSepPunct{\mcitedefaultmidpunct}
{\mcitedefaultendpunct}{\mcitedefaultseppunct}\relax
\EndOfBibitem
\bibitem[Dufferwiel \latin{et~al.}(2018)Dufferwiel, Lyons, Solnyshkov, Trichet,
  Catanzaro, Withers, Malpuech, Smith, Novoselov, Skolnick, Krizhanovskii, and
  Tartakovskii]{Dufferwiel2018}
Dufferwiel,~S.; Lyons,~T.~P.; Solnyshkov,~D.~D.; Trichet,~A. A.~P.;
  Catanzaro,~A.; Withers,~F.; Malpuech,~G.; Smith,~J.~M.; Novoselov,~K.~S.;
  Skolnick,~M.~S.; Krizhanovskii,~D.~N.; Tartakovskii,~A.~I. \emph{Nature
  Communications} \textbf{2018}, \emph{9}, 4797\relax
\mciteBstWouldAddEndPuncttrue
\mciteSetBstMidEndSepPunct{\mcitedefaultmidpunct}
{\mcitedefaultendpunct}{\mcitedefaultseppunct}\relax
\EndOfBibitem
\bibitem[{Fernandez} \latin{et~al.}(2017){Fernandez}, {Russo}, and
  {Barnes}]{Fernandez2017}
{Fernandez},~H.; {Russo},~S.; {Barnes},~W.~L. Design of strong-coupling
  microcavities for optoelectronic applications. 2017 Conference on Lasers and
  Electro-Optics Europe European Quantum Electronics Conference
  (CLEO/Europe-EQEC). 2017; pp 1--1\relax
\mciteBstWouldAddEndPuncttrue
\mciteSetBstMidEndSepPunct{\mcitedefaultmidpunct}
{\mcitedefaultendpunct}{\mcitedefaultseppunct}\relax
\EndOfBibitem
\bibitem[Li \latin{et~al.}(2014)Li, Chernikov, Zhang, Rigosi, Hill, van~der
  Zande, Chenet, Shih, Hone, and Heinz]{Li2014}
Li,~Y.; Chernikov,~A.; Zhang,~X.; Rigosi,~A.; Hill,~H.~M.; van~der
  Zande,~A.~M.; Chenet,~D.~A.; Shih,~E.-M.; Hone,~J.; Heinz,~T.~F. \emph{Phys.
  Rev. B} \textbf{2014}, \emph{90}, 205422\relax
\mciteBstWouldAddEndPuncttrue
\mciteSetBstMidEndSepPunct{\mcitedefaultmidpunct}
{\mcitedefaultendpunct}{\mcitedefaultseppunct}\relax
\EndOfBibitem
\bibitem[Chernikov \latin{et~al.}(2015)Chernikov, van~der Zande, Hill, Rigosi,
  Velauthapillai, Hone, and Heinz]{Chernikov2015}
Chernikov,~A.; van~der Zande,~A.~M.; Hill,~H.~M.; Rigosi,~A.~F.;
  Velauthapillai,~A.; Hone,~J.; Heinz,~T.~F. \emph{Physical Review Letters}
  \textbf{2015}, \emph{115}, 126802\relax
\mciteBstWouldAddEndPuncttrue
\mciteSetBstMidEndSepPunct{\mcitedefaultmidpunct}
{\mcitedefaultendpunct}{\mcitedefaultseppunct}\relax
\EndOfBibitem
\bibitem[Houdr\'e \latin{et~al.}(1995)Houdr\'e, Gibernon, Pellandini, Stanley,
  Oesterle, Weisbuch, O'Gorman, Roycroft, and Ilegems]{Houdre1995}
Houdr\'e,~R.; Gibernon,~J.~L.; Pellandini,~P.; Stanley,~R.~P.; Oesterle,~U.;
  Weisbuch,~C.; O'Gorman,~J.; Roycroft,~B.; Ilegems,~M. \emph{Phys. Rev. B}
  \textbf{1995}, \emph{52}, 7810--7813\relax
\mciteBstWouldAddEndPuncttrue
\mciteSetBstMidEndSepPunct{\mcitedefaultmidpunct}
{\mcitedefaultendpunct}{\mcitedefaultseppunct}\relax
\EndOfBibitem
\bibitem[Gao \latin{et~al.}(2018)Gao, Li, Bamba, and Kono]{Gao2018}
Gao,~W.; Li,~X.; Bamba,~M.; Kono,~J. \emph{Nature Photonics} \textbf{2018},
  \emph{12}, 362--367\relax
\mciteBstWouldAddEndPuncttrue
\mciteSetBstMidEndSepPunct{\mcitedefaultmidpunct}
{\mcitedefaultendpunct}{\mcitedefaultseppunct}\relax
\EndOfBibitem
\bibitem[Chakraborty \latin{et~al.}(2018)Chakraborty, Gu, Sun, Khatoniar,
  Bushati, Boehmke, Koots, and Menon]{Chakraborty2018}
Chakraborty,~B.; Gu,~J.; Sun,~Z.; Khatoniar,~M.; Bushati,~R.; Boehmke,~A.~L.;
  Koots,~R.; Menon,~V.~M. \emph{Nano Letters} \textbf{2018}, \emph{18},
  6455--6460\relax
\mciteBstWouldAddEndPuncttrue
\mciteSetBstMidEndSepPunct{\mcitedefaultmidpunct}
{\mcitedefaultendpunct}{\mcitedefaultseppunct}\relax
\EndOfBibitem
\bibitem[Zhang \latin{et~al.}(2017)Zhang, Feng, Wang, Yang, and
  Wang]{Zhang2017}
Zhang,~K.; Feng,~Y.; Wang,~F.; Yang,~Z.; Wang,~J. \emph{J. Mater. Chem. C}
  \textbf{2017}, \emph{5}, 11992--12022\relax
\mciteBstWouldAddEndPuncttrue
\mciteSetBstMidEndSepPunct{\mcitedefaultmidpunct}
{\mcitedefaultendpunct}{\mcitedefaultseppunct}\relax
\EndOfBibitem
\bibitem[Zhu \latin{et~al.}(2015)Zhu, Chen, and Cui]{Zhu2015}
Zhu,~B.; Chen,~X.; Cui,~X. \emph{Scientific Reports} \textbf{2015}, \emph{5},
  9218\relax
\mciteBstWouldAddEndPuncttrue
\mciteSetBstMidEndSepPunct{\mcitedefaultmidpunct}
{\mcitedefaultendpunct}{\mcitedefaultseppunct}\relax
\EndOfBibitem
\bibitem[Schmitt-Rink \latin{et~al.}(1985)Schmitt-Rink, Chemla, and
  Miller]{Schmitt-Rink1985}
Schmitt-Rink,~S.; Chemla,~D.~S.; Miller,~D.~A. \emph{Physical Review B}
  \textbf{1985}, \emph{32}, 6601--6609\relax
\mciteBstWouldAddEndPuncttrue
\mciteSetBstMidEndSepPunct{\mcitedefaultmidpunct}
{\mcitedefaultendpunct}{\mcitedefaultseppunct}\relax
\EndOfBibitem
\bibitem[Ye \latin{et~al.}(2014)Ye, Cao, O'Brien, Zhu, Yin, Wang, Louie, and
  Zhang]{Ye2014}
Ye,~Z.; Cao,~T.; O'Brien,~K.; Zhu,~H.; Yin,~X.; Wang,~Y.; Louie,~S.~G.;
  Zhang,~X. \emph{Nature} \textbf{2014}, \emph{513}, 214--218\relax
\mciteBstWouldAddEndPuncttrue
\mciteSetBstMidEndSepPunct{\mcitedefaultmidpunct}
{\mcitedefaultendpunct}{\mcitedefaultseppunct}\relax
\EndOfBibitem
\bibitem[Chernikov \latin{et~al.}(2015)Chernikov, Ruppert, Hill, Rigosi, and
  Heinz]{Chernikov2015b}
Chernikov,~A.; Ruppert,~C.; Hill,~H.~M.; Rigosi,~A.~F.; Heinz,~T.~F.
  \emph{Nature Photonics} \textbf{2015}, \emph{9}, 466--470\relax
\mciteBstWouldAddEndPuncttrue
\mciteSetBstMidEndSepPunct{\mcitedefaultmidpunct}
{\mcitedefaultendpunct}{\mcitedefaultseppunct}\relax
\EndOfBibitem
\end{mcitethebibliography}

\end{document}